\documentclass{aa2}
\usepackage{amssymb}
\usepackage{txfonts}
\usepackage{graphicx}
\usepackage{natbib}

\def\beq{\begin{eqnarray}}
\def\eeq{\end{eqnarray}}
\def\eps{\epsilon}

\begin{document}

\title{Revisiting the hot matter in the center of gamma-ray bursts and supernova}
\author{Ang Li\inst{1,2} \and Tong Liu\inst{1,2}}

\institute{Department of Astronomy and Institute of Theoretical
Physics and Astrophysics, Xiamen University, Xiamen, Fujian 361005,
China; \email{liang@xmu.edu.cn (AL),tongliu@xmu.edu.cn (TL)}\and
State Key Laboratory of Theoretical Physics, Institute of
Theoretical Physics, Chinese Academy of Sciences, Beijing 100190,
China}


\titlerunning{The hot matter in GRBs and SNs}
\authorrunning{Li \& Liu}

\date{Received   / Accepted }

\abstract {} {Hot matter with nucleons can be produced in the inner
region of the neutrino-dominated accretion flow in gamma-ray bursts
or during the proto-neutron star birth in successful supernova. The
composition and equation of state of the matter depend on the
dynamic $\beta$ equilibrium under various neutrino opacities. The
strong interaction between nucleons may also play an important role.
We plan to extend the previous studies by incorporating these two
aspects in our model.} {The modification of the $\beta$-equilibrium
condition from neutrino optically thin to thick has been modeled by
an equilibrium factor $\chi$ ranging between the
neutrino-freely-escaping case and the neutrino-trapped case. We
employ the microscopic Brueckner-Hartree-Fock approach extended to
the finite temperature regime to study the interacting nucleons.}
{We show that the composition and chemical potentials of the hot
nuclear matter for different densities and temperatures at each
stage of $\beta$ equilibrium. We also compare our realistic equation
of states with those of the free gas model. We find the neutrino
opacity and the strong interaction between nucleons are important
for the description and should be taken into account in model
calculations.} {}

 \maketitle

\section{Introduction}

Gamma-ray bursts (GRBs) and supernova (SNs) are extremely powerful
explosions in the universe. In the centers of these objects, hot
dense matter may be generated. The properties of the matter, such as
its composition and equation of state (EoS), are very important for
the studies of GRBs and SNs.

First, for the central engines of GRBs, one of the plausible
candidates is the neutrino-dominated accretion flow (NDAF) around a
rotating stellar-mass black hole. Such systems may originate from
the merger of two compact objects or a collapsar. The NDAF model has
been widely applied to explain the energy source and several
observations of GRBs in the past decade
\citep[e.g.,][]{Popham1999,Narayan2001,Kohri2002,Di
Matteo2002,Kohri2005,Lee2005,Gu2006,Chen2007,Janiuk2007,Kawanaka2007,Lei2009,Liu2007,Liu2008,Liu2010a,Liu2010b,
Liu2012a,Liu2012b,Liu2013,Sun2012,Kawanaka2012,Kawanaka2013}. Hot
matter approaching nuclear densities ($\rho\sim \rm 10^{10} -10^{13}
g~cm^{-3}$ and $T \sim \rm 10^{10}-10^{11} K$) may appear in the
inner regions of those disks, so a proper description for them
should be included in the NDAF model. \citet{Liu2007} studied the
radial structure and the neutrino annihilation luminosity of the
NDAF. They introduced arbitrarily a bridging formula to treat the
radial distribution of the electron fraction between neutrino
optically thin and thick limits, thanks to an analytical relation of
the chemical potential equilibrium obtained for the former case by
\citet{Yuan2005}. But they ignored the strong interaction between
nucleons, and simplified the calculations by suggesting that the
matter was in a free-gas state, which essentially meant that there
was no difference between the number density and the energy density.
\citet{Kawanaka2007} paid attention to the difference of the number
density and the energy density, but an active connection between
neutrino optically thin and thick limits was not included in their
calculations. Therefore, a more improved theoretical NDAF model
should be presented, especially for the inner region of a disk.

Second, SNs (or collapsars) have also been widely calculated or
simulated in the past decade \citep[see,
e.g.,][]{MacFadyen2001,Proga2003,Buras2006,Burrows2007,Iwakami2008,Hammer2010}.
In a successful SN, the birth of a proto-neutron star (PNS) may
mainly go through several distinct steps \citep[see,
e.g.,][]{Prakash2001}. The first step lasts less than one minute,
during which the star with a neutrino-trapped core of mass
experiences the core bounce and the passage of a shock through the
star's mantle. The outer mantle is both accreting matter from the
surrounding area and losing energy due to the thermal neutrino
emission. In the second step, the accretion is no longer important
and neutrino cooling dominates. During the above two steps, a hot
dense state ($\rho\sim \rm 10^{12} -10^{14} g~cm^{-3}$ and $T \sim
\rm 10^{10} K$) is present both in the core and in the outer part.
In such a state the neutrino production exists along with their
transportation, hence the dynamics of the production process should
play an important role in the calculation and simulation of SNs,
where the opacity has to be taken into account.

Finally, because neutrino radiation is the main cooling mechanism
for the hot matter of GRBs and SNs, the chemical potential
equilibrium in the matter may depend on the neutrino opacity, namely
the dynamics of the $\beta$ processes~\citep{Imshennik1967}. Also,
nucleons may interact with each other in such dense matter, thus an
improved EoS including the strong interaction becomes imperative. In
this paper, we then focus on the effects of the dynamics of the
$\beta$ processes and the strong interaction on various input
microphysics of GRBs and SNs, such as the relative composition and
the EoS of the matter.

Accordingly, we assume that the nuclei are dissolved completely into
nucleons, therefore nuclear many-body theories are applicable for
the deriving of nucleonic chemical potentials. Incorporating the
strong interaction between nucleons will certainly affect the
constituent chemical potentials, the composition and the EoS of the
matter. The employed nuclear model is the microscopic
Brueckner-Hartree-Fock (BHF) approach widely used for the study of
dense stellar matter and neutron star properties
~\citep{bbb,book,baldo,bhf1,liang4,liang5,bhf3,liang1,liang2,liang3,bhf2},
as we shall discuss in section 2.2. We stress that our aim is not to
model the centers of GRBs and SNs; rather, we want to explore how
the properties of the hot matter in the center depend on the
strong-interaction effect and the dynamic $\beta$-process related to
neutrinos. Such effects are usually missing in most of the GRBs' and
SNs' studies.

The paper is organized as follows. In section 2, we establish our
physical model and describe in details the numerical methods for the
calculation. In section 3, numerical results are presented. We
present our main conclusions in section 4.

\section{Model}
\subsection{$\beta$ equilibrium of the hot nuclear matter}

For the hot matter where neutrino are completely trapped, the system
can achieve its equilibrium via the following $\beta$ processes,
\beq
&& e^-+p \rightleftharpoons  n+ \nu_e\:, \label{eq:e_capR}\\
&& e^++n \rightleftharpoons  p+ \bar{\nu}_e\:, \label{eq:p_capR}\\
&& n     \rightleftharpoons  p+e^- + \bar{\nu}_e\:.
\label{eq:n_decR} \eeq The reaction rates of the $\beta$ reactions
are equal to those of the corresponding inverse processes. Because
the photons are also trapped (the chemical potential of photons
$\mu_{\gamma}=0$), the chemical equilibria $\gamma + \gamma
\rightleftharpoons e^+ + e^- \rightleftharpoons \nu_{e} +
\bar{\nu}_{e} $ give \beq
 \mu_{e^-} = -\mu_{e^+} \:.\eeq
Then by writing $\mu_{e^-} \equiv \mu_e$, the usual chemical
equilibrium condition can be applied as \beq \mu_n + \mu_{\nu_e}=
\mu_p+\mu_e \:,\label{eq:trapping20} \eeq or \beq \mu_n =
\mu_p+\mu_e \:, \label{eq:trapping2} \eeq if the chemical potential
of the trapped neutrinos is zero. These are the well-known chemical
equilibrium conditions which are generally used to determine the
composition of the hot matter under $\beta$ equilibrium
\citep{bhf3,liang1,liang2,liang3,bhf2}.

However, if neutrinos can leave the system freely, the
$\beta$-equilibrium of the $\beta$ reactions cannot be treated as a
chemical equilibrium problem. In such circumstance the steady state
of the hot matter is achieved under the following condition: \beq
\lambda_{e^-p} =\lambda_{e^+n} +\lambda_{n}\:,
    \label{eq:equil}
\eeq where $\lambda_{e^-p}, \lambda_{e^+n}$ stands for the rate of
the $e^{\pm}$-captures, \beq
e^-+p &\rightarrow&  n+ \nu_e\:, \label{eq:e_cap}\\
e^++n &\rightarrow&  p+ \bar{\nu}_e \:, \label{eq:p_cap} \eeq and
$\lambda_{n}$ is the rate of neutron decay,
 \beq n \rightarrow  p+e^- + \bar{\nu}_e \:. \label{eq:n_dec} \eeq
The reaction rates $\lambda$ should be functions of the temperature
$T$ and the constituent chemical potentials. Compared with the rate
of positron capture by neutrons $\lambda_{e^+n}$, the rate of
neutron decay $\lambda_{n}$ in this case could be neglected,
therefore the $\beta$-equilibrium condition is reduced to
$\lambda_{e^-p}=\lambda_{e^+n}$ \citep{yuan00,yuan0,Yuan2005}, and
finally \beq \mu_n=\mu_p+2\mu_e \label{eq:nd_eq}\:,  \eeq after the
used assumption of the elastic approximation
\citep[e.g.,][]{Yuan2005}. Again zero chemical potential is applied
for neutrinos.

By combining Eqs.~(\ref{eq:trapping2}) and (\ref{eq:nd_eq}) we
introduce an equilibrium factor $\chi$ in the range of $[1,~2]$, to
explore the effect of the dynamics of the $\beta$ processes, namely
\beq \mu_n = \mu_p+\chi \mu_e \:, \label{eq:trapping} \eeq where
$\chi = 1$ suggests completely neutrino-trapped matter, and $1<\chi
\leqslant 2$ corresponds to the matter with a certain amount of
freely-escaping neutrinos.

To find the composition and the EoS of the matter, for each baryon
number density $n_{\rm B}$, Eq.~(\ref{eq:trapping}) should be solved
together with the charge neutrality condition, \beq
n_{e^-}-n_{e^+}=n_{p} \:, \label{e:neutral} \eeq and the
conservation of the baryon number, \beq n_n+n_p = n_{\rm B}\:.
    \label{eq:baryon} \eeq

The nucleonic chemical potentials are derived from the free energy
density of the nuclear matter, based on the finite-temperature BHF
nuclear many-body approach (as illustrated below in section 2.2).
The chemical potentials of the non-interacting leptons $e^{\pm}$ are
obtained by solving numerically the free Fermi gas model at a finite
temperature. Specifically, the number density of species
$i~(i=e^{\pm})$ is written as \beq {n_i}(T) = \int_0^{\infty}
\frac{g}{(2 \pi \hbar)^3} n^F_i(k,T) d^3k \:, \eeq where $(2\pi
\hbar)^3$ is the "unit" volume of a cell in the phase space and $g$
is the number of states of a particle with a given momentum $k$. For
$e^{\pm}$ leptons, $g$ equals 2. Finally, $n^F_i(k,T)$ is the
Fermi--Dirac--Statistic, \beq n^F_{e^-}(k,T) & = & \frac{1} {{\rm
exp}[(E(k) - \mu_e)/k_BT]+1},
\\n^F_{e^+}(k,T) & = & \frac{1} {{\rm exp}[(E(k) + \mu_e)/k_BT]+1} \:,\eeq
where $k_B$ is the Boltzmann constant and the energy $E(k) =
\sqrt{k^2+m_e^2}$ with $m_e$ being the mass of the electrons or
positrons.

Once the nucleonic and leptonic chemical potentials are determined,
one can proceed to calculate the composition of the hot
$\beta$-equilibrium matter by solving Eqs. (\ref{e:neutral}) and
(\ref{eq:baryon}) together with Eq.~(\ref{eq:trapping}). Then the
total energy density $\rho$ and the total pressure $P$ of the system
are:\beq
 \rho &=& \rho_{l} + \rho_{\rm B}\:,  \eeq \beq
 P &=& P_{l} + P_{\rm B}\:,
\eeq where $\rho_l$ and $P_l$ are the standard contributions of the
leptons: \beq
 \rho_{l}  & = & \frac{8 \pi}{(2 \pi \hbar)^3}
 \int_0^{\infty}(f_{e^-}-f_{e^+})  E(k) k^2 d k\:,
   \\
 P_{l} & = &\ \frac{8 \pi}{3(2 \pi \hbar)^3}
 \int_0^{\infty}(f_{e^-}-f_{e^+})  k^4 /E(k)d k\:.
 \label{eq:electroneps} \
\eeq We present in the following subsection how the contributions of
baryons ($\rho_{\rm B},~P_{\rm B}$) are determined.

\subsection{BHF nuclear many-body approach}

Currently, one of the most advanced microscopic approaches to the
EoS of the nuclear matter is the BHF model \citep{book}. Recently,
this model was extended to the finite-temperature regime within the
Bloch-De Dominicis formalism \citep{bloch1,bloch2,bloch3}. The
central quantity of the BHF formalism is the $G$-matrix, which in
the finite-temperature extension
\citep{bloch1,bloch2,bloch3,book,baldo} is determined by solving
numerically the Bethe-Goldstone equation, and can be written in
operatorial form as \beq
  G_{ab}[W] = V_{ab} +
  \sum_c \sum_{p,p'}
  V_{ac} \big|pp'\big\rangle
  { Q_c \over W - E_c +i\eps}
  \big\langle pp'\big| G_{cb}[W]\:,
\label{e:g} \eeq where the indices $a,b,c$ indicate pairs of
nucleons and the Pauli operator $Q$ and energy $E$ determine the
propagation of intermediate nucleon pairs. In a given
nucleon-nucleon channel $c=(12)$ one has \beq
   Q_{(12)} &=& [1-n^F_1(k_1)][1-n^F_2(k_2)]\:,
\eeq \beq
  E_{(12)} &=& m_1 + m_2 + e_1(k_1) + e_2(k_2)\:,
\label{e:e} \eeq with the single-particle (s.p.) energy $e_i(k) =
k^2\!/2m_i + U_i(k)$, the above-mentioned Fermi distribution
$n^F_i(k)=\big( e^{[e_i(k) - \tilde{\mu_i}]/T} + 1 \big)^{-1}$, the
starting energy $W$, and the two-body interaction (bare potential)
$V$ as fundamental input. The various s.p. potentials within the
continuous choice are given by \beq
  U_1(k_1) = {\rm Re}\!\!\!\!
  \sum_{2=n,p}\sum_{k_2} n(k_2)
  \big\langle k_1 k_2 \big| G_{(12)(12)}\left[E_{(12)}\right]
  \big| k_1 k_2 \big\rangle_A\:,
\label{e:u} \eeq where $k_i$ generally denote momentum and spin. For
given partial densities $n_i\; (i=n,p)$ and temperature $T$,
Eqs.~(\ref{e:g}-\ref{e:u}) have to be solved self-consistently along
with the equations for the auxiliary chemical potentials
$\tilde{\mu_i}$, $ n_i = \int_k n^F_i(k) $.

Regarding the interactions, we use the Argonne $V_{18}$
nucleon-nucleon potential \citep{v18} together with the microscopic
nuclear three-body forces (TBF) \citep{tbf1,tbf2,tbf3}. The
including of TBF accomplishes excellently two important tasks.
First, the corresponding zero-temperature nuclear EoS reproduces the
nuclear matter saturation point correctly and fulfills several
requirements from the nuclear phenomenology \citep{bbb}. Second, the
main relativistic effect can be taken into account, and the results
in our nonrelativistic scheme agree well with the predictions of the
corresponding relativistic approaches \citep{tbf3}.

Once the different s.p.~potentials for the species $i=n,~p$ are
known, the free energy density of nuclear matter has the following
simplified expression \beq
 f = \sum_i \left[ \sum_{k} n^F_i(k)
 \left( {k^2\over 2m_i} + {1\over 2}U_i(k) \right) - Ts_i \right]\:,
\label{e:f} \eeq where \beq
 s_i = - \sum_{k} \Big( n^F_i(k) \ln n^F_i(k) + [1-n^F_i(k)] \ln [1-n^F_i(k)] \Big)
\eeq is the entropy density for component $i$ treated as a free gas
with s.p.~spectrum $e_i(k)$ \citep{book,baldo}.

All thermodynamic quantities of interest can then be computed from
the free energy density, Eq.~(\ref{e:f}); namely, the ``true"
chemical potentials $\mu_i\; (i=n,~p)$, internal energy density
$\rho_{\rm B}$, and pressure $P_{\rm B}$ are \beq
 \mu_i &=& {{\partial f}\over{\partial n_i}} \:,
\\
s &=& -{{\partial f}\over{\partial T}} \:,
\\
 \rho_{\rm B} &=& f + Ts \:,
 \\
  P_{\rm B} &=& n^2 {\partial{(f/n)}\over \partial{n}}
 = \sum_i \mu_i n_i - f\:.
\label{e:eps} \eeq

\section{Results}

\begin{figure}
\centering
\includegraphics[width=0.45\textwidth]{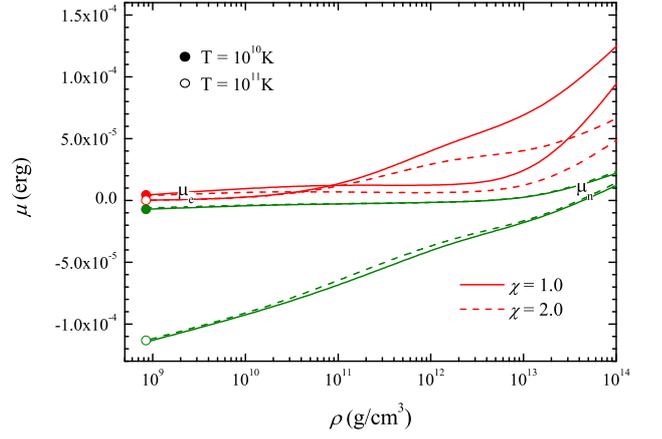}
\caption{Chemical potentials of both electrons and neutrons (red
lines for $\mu_e$, and green ones for $\mu_n$) as a function of the
energy density $\rho$ at two fixed temperatures $T = 10^{10} \rm K$
(filled symbol) and $T = 10^{11} \rm K$ (open symbol), for both
$\chi = 1$ (solid lines) and $\chi = 2$ (dashed lines) cases,
respectively.} \label{fig1}
\end{figure}

For both GRBs and SNs, a hot state with nucleons exists with
$\rho\sim \rm 10^{9} -10^{14} g~cm^{-3}$ and $T \sim \rm 10^{9}
-10^{11} K $. We adopted these parameter ranges in our model. One
additional parameter $\chi$, the so-called equilibrium factor, is
introduced to incorporate the effect of the neutrino opacity, and
its value should be between $1$ and $2$, as discussed in the
previous section.

\begin{figure}
\centering
\includegraphics[width=0.45\textwidth]{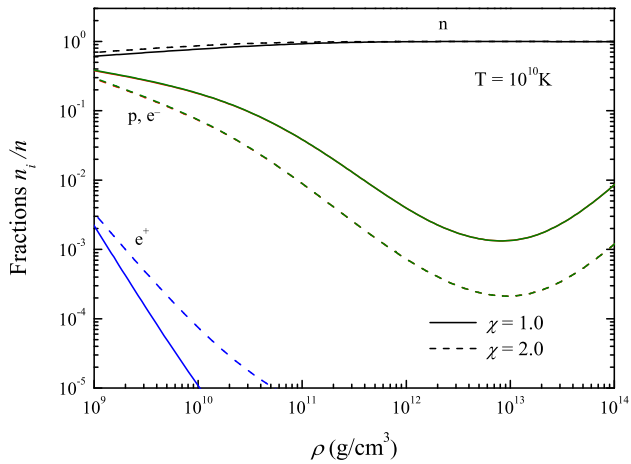}
\caption{Relative fractions $n_i/n_B~(i=n,~p,~e^-,~e^+)$ as a
function of the energy density $\rho$ at fixed temperature $T =
10^{10} \rm K$, for both $\chi = 1$ (solid lines) and $\chi = 2$
(dashed lines) cases.} \label{fig2}
\end{figure}

\begin{figure}
\centering
\includegraphics[width=0.45\textwidth]{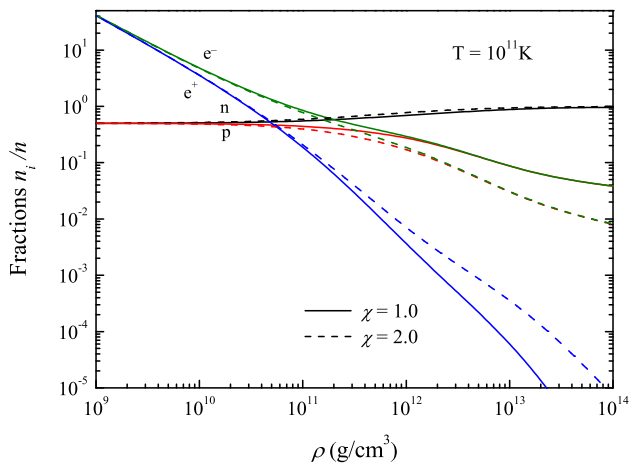}
\caption{Same with Fig.~\ref{fig2}, but for $T = 10^{11} \rm K$.}
\label{fig3}
\end{figure}
We first display in Fig.~\ref{fig1} the chemical potentials of both
electrons and neutrons (red lines for $\mu_e$, and green ones for
$\mu_n$) as a function of the energy density $\rho$ at two fixed
temperatures $T = 10^{10} \rm K$ (filled symbol) and $T = 10^{11}
\rm K$ (open symbol), for both $\chi = 1$ (solid lines) and $\chi =
2$ (dashed lines) cases. Regardless of the temperature, the electron
chemical potentials are always positive and increase monotonously
with the density. Since electrons are treated as a degenerate fermi
gas, the corresponding degenerate pressures are expected to increase
with the density as well. However, $\mu_n$ changes its sign from
negative to positive at high densities (around $\rm 10^{14}
g~cm^{-3} $), which simply means that the strong interaction
dominates for such dense matter. We mention here the nuclear
saturation density is about $\rm 2.5 \times 10^{14} g~cm^{-3}$.
Moreover, we find that the equilibrium factor $\chi$ affects $\mu_n$
only slightly. Its influence on $\mu_e$ is evident, but mainly at
the high-density region. With the increase of $\chi$ parameter, the
electron chemical potential $\mu_e$ is largely reduced, which means
there is a reduced lepton fraction in the matter. Also, compared
with the low temperature ($T = 10^{10} \rm K$) case, we find much
rapid increases of both $\mu_e$ and $\mu_n$ at high temperature ($T
= 10^{11} \rm K$) case.

Those results should have significant impacts on the study of the
NDAF model, since we know that, there is a wide range of density in
the inner region of the disk \citep[see, e.g.,][]{Liu2007,Liu2008},
$\rho \sim \rm 10^9-10^{13} g~cm^{-3}$, and for a typical accretion
rate of $1M_\odot~\rm s^{-1}$, the neutrino optical depth would
change from thick to thin if one moves away from the central black
hole. Therefore, one should include the dependence of the
constituent chemical potentials on the equilibrium condition, the
temperature and the energy density, as demonstrated in the above
figure.

\begin{figure}
\centering
\includegraphics[width=0.45\textwidth]{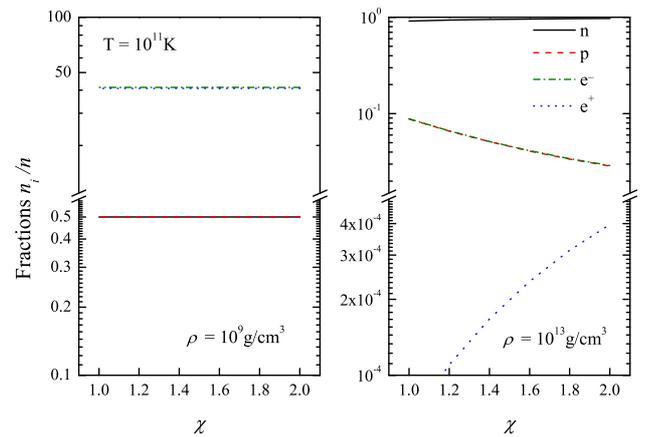}
\caption{Relative fractions $n_i/n_B~(i=n,~p,~e^-,~e^+)$ as a
function of $\chi$ parameter at two fixed densities $\rho =\rm
10^{9} g ~cm^{-3}$ (left panel) and $\rm 10^{13} g~cm^{-3}$ (right
panel) for $T = 10^{11} \rm K$.} \label{fig4}
\end{figure}
We then show the compositions of the matter in Fig.~\ref{fig2} (for
$T = 10^{10} \rm K$) and Fig.~\ref{fig3} (for $T = 10^{11} \rm K$).
The relative fractions $n_i/n_B~(i=n,~p,~e^-,~e^+)$ are plotted as a
function of the energy density $\rho$, for both $\chi = 1$ (solid
lines) and $\chi = 2$ (dashed lines) cases. In the low temperature
case of $T = 10^{10} \rm K$ (Fig.~\ref{fig2}), the proton fractions
are very similar to the electron fractions, and they both increase
with the density as a combined result of an increased electron
chemical potential $\mu_e$ and the charge neutrality condition (Eq.
\ref{e:neutral}). On the contrary, the positron fractions decrease
very quickly with increasing density as a natural result of the
Fermi-Dirac distribution employed for the leptons. Here neutrons
compose most of the matter (larger than $\sim 70\%$ ), and the
matter is practically $npe^-$ for the chosen temperature $T =
10^{10} \rm K $. If the matter is much hotter, for example $T =
10^{11} \rm K $ in Fig.~\ref{fig3}, at relatively low density (below
$\rm 10^{11} g~cm^{-3}$), a large amount of $e^{\pm }$ pairs exist
regardless the choice of $\chi$ parameter. This means the efficient
creation of the $e^{\pm }$ pair is the characteristic of a hot
nuclear system.

Fig.~\ref{fig4} shows the compositions as a function of $\chi$
parameter at two fixed densities $\rho =\rm 10^{9} g ~cm^{-3}$ (left
panel) and $\rm 10^{13} g~cm^{-3}$ (right panel) for a fixed
temperature of $T = 10^{11} \rm K$.  At low densities around $\rm
10^{9} g ~cm^{-3}$, the variation of the equilibrium parameter
$\chi$ affects only trivially the relative fractions, as one might
also notice in Fig.~\ref{fig3}. But at high densities the $\chi$
dependence becomes important, thus the accretion matter in the inner
region of the disk should be more appropriately modeled in the NDAF
model.

\begin{figure}
\centering
\includegraphics[width=0.45\textwidth]{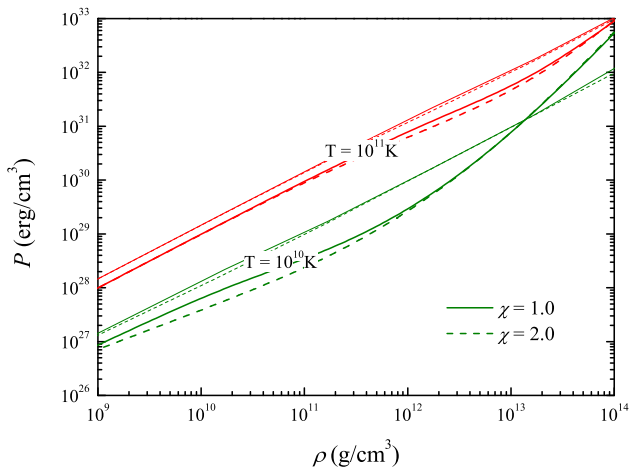}
\caption{EoSs for two temperatures $T = 10^{10} \rm K$ (red lines)
and $T = 10^{11} \rm K$ (green lines), and for both $\chi = 1$
(solid lines) and $\chi = 2$ (dashed lines) cases. The predictions
of free gas model are also shown in thin lines for comparison.}
\label{fig5}
\end{figure}

Finally we summarize our results of EoSs in Fig.~\ref{fig5} for two
temperatures $T = 10^{10} \rm K$ (red lines) and $T = 10^{11} \rm K$
(green lines), and for both $\chi = 1$ (solid lines) and $\chi = 2$
(dashed lines) cases. The predictions of the free gas model are also
shown in thin lines for comparison. In both models, the higher
temperature, the stiffer EoS. Because electrons contribute equally
to our model and the free gas model, the differences between the
thick and thin lines arise from the difference in the nucleon part.
Our realistic EoSs (thick lines) usually lie below the ones based on
the free-gas approximation (thins lines) because it is the
long-range gravity who dominates in the matter, not the short-range
nucleon-nucleon force, except at high densities. The strong
interaction can not be ignored in such dense matter.

As commonly recognized, the pressure from nucleons should dominate
in the inner region of the NDAF when the mass accretion rate is
larger than $0.01M_\odot~\rm s^{-1}$ \citep[see,
e.g.,][]{Chen2007,Liu2007,Kawanaka2007}. As shown above, the
nucleonic EoS is subject to change if we include more microscopic
physics beyond the simple standard free gas model. Such simple model
is widely used in the NDAF or collapsar models \citep[see,
e.g.,][]{Popham1999,MacFadyen2001,Chen2007,Liu2007}. For an improved
study, a detailed database of the resulting pressure with the change
of the temperature, the density, and the neutrino opacity should be
built.

\section{Conclusion}

In this paper, various properties of the hot nuclear matter possible
in the inner regions of GRBs and SNs have been revisited. We employ
the microscopic BHF approach to account for the strong interaction
between nucleons, and calculate the nucleonic chemical potentials
and the nucleonic EoS in this method. We introduce a parameterized
chemical potential equilibrium bridging between neutrino optically
thin and thick limits, and show the compositions and the EoSs of the
matter at several temperatures under different chemical potential
equilibria, labeled by the so-called equilibrium factor $\chi$. We
also display the comparison of the EoSs with those of the free gas
model. We find that, for the description of the hot matter the
effect of the neutrino opacity and the strong interaction can be
very important. They should be taken into account in future model
calculations.

For example, since one popular central engine model that powers GRBs
consists of a black hole and a NDAF with a hypercritical mass
accretion rate, and the neutrino annihilation luminosity in the NDAF
model can be significantly affected by the matter properties of the
inner region of the disk, we will revisit the NDAF model by
incorporating the findings of the present work in the description of
the inner region, to verify more convincingly whether the NDAF model
still can be one of the candidates of the central engines of GRBs.

\begin{acknowledgements}
We would like to thank Prof. T. T. Fang who carefully read the
manuscript and made valuable suggestions. This work was supported by
the National Basic Research Program (973 Program) of China under
Grant 2009CB824800 and the National Natural Science Foundation of
China under grants 11103015 and 11233006.
\end{acknowledgements}

\clearpage

\end{document}